\documentclass[12pt]{article}
\usepackage{hfstyle}
\usepackage{graphicx}
\newtheorem {thm}{Theorem}[section]
\newtheorem {lem}[thm]{Lemma}
\newtheorem {cor}[thm]{Corollary}

\def\bn{\begin{eqnarray*}}
\def\en{\end{eqnarray*}}
\def\bnn{\begin{eqnarray}}
\def\enn{\end{eqnarray}}

\def\E{{\sf{E\,}}}

\def\P{{\sf{P}}}

\def\ttau{\tilde{T}}
\def\r{{\bf r}}
\def\de{{d_{PE}}}

\def\R{{\sf{R}}}

\def\L{{\mathcal{L}^p}}
\def\tL{\tilde{\cal{L}}}
\def\Cox{\hfill \mbox{\sl Q.E.D.}}

\def\|{\, | \, }
\def\ee{\varepsilon}

\begin{document}
\title{Packet delay in models of data networks}
\author{Henryk Fuk\'s$^{1,3}$, Anna T. Lawniczak$^{2,3}$ and Stanislav Volkov$^{4}$
      \oneaddress{
         ${^1}$Department of Mathematics \\ Brock University \\
         St. Catharines, Ontario L2S 3A1, Canada\\
         {~}\\
         ${^2}$Department of Mathematics and
         Statistics, \\University of Guelph, \\
         Guelph, Ontario N1G 2W1, Canada\\
         {~}\\
         ${^3}$The Fields Institute for Research\\
         in Mathematical Sciences\\
         Toronto, Ontario M5T 3J1, Canada\\
         {~}\\
         ${^4}$De Technische Universiteit Eindhoven\\
         LG 1.1.9 TUE - EURANDOM\\
         P.O.Box  513-5600 MB Eindhoven\\
         The Netherlands\\
         {~}\\
         \email{hfuks@brocku.ca,\\
         alawnicz@fields.utoronto.ca,\\
         svolkov@euridice.tue.nl}
       }
   }

\Abstract{We investigate individual packet delay in a model of data networks
with table-free, partial table and full table routing. We present
analytical estimation for the average packet delay in a network with
small partial routing table. Dependence of the delay on the size of
the network and on the size of the partial routing table is examined
numerically. Consequences for network scalability are discussed.}
\maketitle

\section{Introduction}

Importance of packet-switched data networks in contemporary society
cannot be overestimated. In an attempt to understand their complex
dynamics, several simplified models have been proposed in recent years
\cite{Campos95,Deane96,paper9,Kadirire94,Ohira98,Tretyakov98}.
The construction of these models have been inspired by successful
and well established in physics methodologies of particle systems,
cellular automata and lattice gas cellular automata. The application
of these methodologies in the context of data networks provides
a promising alternative approach. Even though some of these models
are simplistic, they can be  expanded and modified to incorporate
various realistic aspects of data networks. Additionally, these models
are not only amenable to computer simulations but also to obtaining
analytical results.

One of the interesting questions which needs to be addressed in the context
of these models is an issue of influence of the randomness present in
the routing algorithm on the network's dynamics and its effects on the
performance of the network.

In \cite{paper9} we investigated a model in which packets are routed
according to a table stored locally at each node. If the table
includes all other nodes of the network, such an algorithm is called
full table routing algorithm. However, if only nodes closer than $m$
links away are present in the table (partial table routing), packets
with a destination address not present in the table are forwarded to a
randomly selected nearest neighbour node. This introduces certain
amount of randomness or noise into the system, and as a result, the
\emph{delay} changes. By delay we mean the time required for a packet
to reach its destination.

In this work, we will investigate how the delay experienced
by a single packet, when no other packets are present,
depends on the degree of randomness in the routing scheme.
While interactions with other packets will obviously strongly
influence the delay, in \cite{paper9} we found that
the delay experienced by a single packet is an important
parameter characterizing the network. For example, simulation
experiments reported in \cite{paper9} seem to indicate that
in many cases the critical load is inversely proportional
to the single packed delay.
In an attempt to gain some insight into properties of this important parameter,
we will derive analytical estimates for the single packet delay
and compare it with direct simulations. Finally, we will discuss how
these results affect scalability of the proposed network model.

\section{Network Models Definitions}

Detailed description of the network model is given in \cite{paper9}. Here,
we summarize only its main features.
The purpose of the network is to transmit messages from points of their origin
to their destination points. In our model, we will assume that the entire
message is contained in a single ``capsule'' of information, which, by
analogy to packet-switching networks, will be simply called a \emph{packet}.
In a real packet-switching network, a single packet carries the information
``payload'', and some additional information related to the internal
structure of the network. We will ignore the information ``payload''
entirely, and assume that the packet carries only two pieces of information:
time of its creation and the destination address.

Our simulated network consists of a number of interconnected nodes. Each
node can perform two functions: of a \emph{host}, meaning that it can
generate and receive messages, and of a \emph{router} (message processor),
meaning that it can store and forward messages. Packets are created and
moved according to a discrete time parallel algorithm. The structure of the
considered networks and the update algorithm will be described in
subsections which follow.

\subsection{Connection Topology}

In this paper, we will consider a connection topology in a
form of a two-dimensional square lattice  with periodic boundary
conditions $\L$. The network hosts and routers are located at nodes of
the lattice $\mathcal{L}^p$. The position of each node on a lattice
$\mathcal{L}^p$ is described by a discrete space variable
$\mathbf{r}$, such that
\begin{equation}
\mathbf{r}=i\mathbf{c}_{x}+j\mathbf{c}_{y},
\end{equation}
where $\mathbf{c}_{x},\mathbf{c}_{y}$ are Cartesian unit vectors, and
$i,j=1,\ldots ,L$. The value of $L$ gives a number of nodes in
the horizontal and vertical direction of the lattice $\mathcal{L}^p$.
We denoted by $C(\mathbf{r})$ the set of all nodes directly connected
with a node $\mathbf{r}$. For each $\mathbf{r}\in $
$
\mathcal{L}^{p}$, the set $C(\mathbf{r})$ is of the form
\begin{equation}
C(\mathbf{r})=\{\mathbf{r}-\mathbf{c}_{x},\mathbf{r}+\mathbf{c}_{x},\mathbf{r
}-\mathbf{c}_{y},\mathbf{r}+\mathbf{c}_{y}\}.  \label{nbh}
\end{equation}
In this case, the node $\mathbf{r}$ is connected with its four nearest
neighbours. In the networks considered here, each node maintains a
queue of unlimited length where the arriving packets are stored.
Packets stored in queues, at individual lattice nodes, must be
delivered to their destination addresses. To assess how far a given
packet is from its destination, we introduce the concept of distance
between nodes. We will use periodic ``Manhattan'' metric to compute
the distance between two nodes ${\mathbf{r}}_{1}=(i_{1},j_{1})$ and ${
\mathbf{r}}_{2}=(i_{2},j_{2})$:
\begin{equation}
d_{PM}({\mathbf{r}}_{1},{\mathbf{r}}_{2})=L-\left|
|i_{2}-i_{1}|-\frac{L}{2}
\right| -\left| |j_{2}-j_{1}|-\frac{L}{2}\right|.
\end{equation}

\subsection{Update Algorithms}

The dynamics of the networks are governed by the parallel update
algorithms similar to the algorithm used in~\cite{Ohira98}. We start
with an empty queue at each node, and with discrete time clock $k$ set
to zero. Then, the following actions are performed in sequence:

\begin{enumerate}
\item  At each node,\ independently of the others, a packet is created with
probability $\lambda$. Its destination address is randomly selected
with uniform probability
distribution among all other nodes in the network. The newly created packet is placed at the end of the
queue.

\item  At each node, one packet (or none, if the local queue is empty) is
picked up from the top of the queue and forwarded to one of its
neighboring sites according to a one of the routing algorithms to be
described below. Upon arrival, the packet is placed at the end of the
appropriate queue. If several packets arrive to a given node at the
same time, then they are placed at the end of the queue in a random
order. When a packet arrives to its destination node, it is
immediately destroyed.

\item  $k$ is incremented by 1.
\end{enumerate}

This sequence of events, which constitutes a \emph{single time step
update}, is then repeated arbitrary number of times. The state of the
network is observed after sub-step 3, before clock increase and repetition
of sub-step 1. In order to explain the routing
algorithms mentioned in sub-step 2, we will first describe one of its
simplified versions.

Let us assume that we measure distance using metric $d_{PM}$. To
decide where to forward a packet located at a node $\mathbf{r}$ with
the destination address $\mathbf{r}_{d}$, two steps are performed:

\begin{enumerate}
\item  From sites directly connected to $\mathbf{r}$, we select sites which
are closest to the destination $\mathbf{r}_{d}$ of the packet. More
formally, we construct a set $A_{\infty }(\mathbf{r})$ such that
\begin{equation}
A_{\infty }(\mathbf{r})=\{\mathbf{a}\in
C(\mathbf{r}):d(\mathbf{a},\mathbf{r}%
_{d})=\min_{\mathbf{x}\in C(\mathbf{r})}d_{PM}(\mathbf{x},\mathbf{r}_{d})\}
\label{seta}
\end{equation}

\item  From $A_{\infty }(\mathbf{r})$, we select a site which has the
smallest queue size. If there are several such sites, then we select one of
them randomly with uniform probability distribution. The packet is forwarded
to this site. Using a formal notation again, we could say that the packet is
forwarded to a site selected randomly and uniformly from elements of a set $%
B_{\infty }(\mathbf{r})$ defined as
\begin{equation}
B_{\infty }(\mathbf{r})=\{\mathbf{a}\in A_{\infty }(\mathbf{r}):n(\mathbf{a}%
,k)=\min_{\mathbf{x}\in A_{\infty }(\mathbf{r})}n(\mathbf{x},k)\},
\end{equation}
where $n(\mathbf{x},k)$ is a queue size at a node $\mathbf{x}$ at time $k$.
\end{enumerate}

To summarize, the routing algorithm $\mathbf{R}_{\infty }$ described
above sends the packet to a site which is closest to the destination
(in the sense of the metric $d_{PM}$), and if there are several such
sites, then it selects from them the one with the smallest queue. If
there is still more than one such node, random selection takes place.
It is clear that each packet routed according to the algorithm
$\mathbf{R}_{\infty }$ will travel to its destination along the
shortest possible path (shortest in the sense of the metric $d_{PM}$,
not necessarily in terms of a number of time steps required to reach
the destination). In real networks, this does not always happen. In
order to allow packets to take alternative routes, not necessarily
shortest path routes, we will introduce a small modification to the
routing algorithm $\mathbf{R}_{\infty }$ described above.

The modified algorithm $\mathbf{R}_{m}$, for each node $\mathbf{r,}$
will use instead of the set $A_{\infty }(\mathbf{r})$ a set
$A_{m}(\mathbf{r})$ defined as follows. In the construction of the set
$A_{m}(\mathbf{r})$ instead of minimizing distance
$d_{PM}(\mathbf{x},\mathbf{r}_{d})$
from $\mathbf{x}$ to the destination $\mathbf{r}_{d}$,
as it was done in $(\ref{seta})$, we will minimize $\Theta _{m}
(d_{PM}(\mathbf{x},\mathbf{r}_{d}))$, where
\begin{equation}
\Theta _{m}(y)=\left\{
\begin{array}{ll}
y, & \mbox{if $y < m$}, \\
m, & \mbox{otherwise},
\end{array}
\right.
\end{equation}
for a given integer $m$. Thus, the definition of the set $A_{m}(\mathbf{r})$
is
\begin{equation}
A_{m}(\mathbf{r})=\{\mathbf{a}\in C(\mathbf{r}):\Theta
_{m}(d_{PM}(\mathbf{a},\mathbf{r}_{d}))=\min_{\mathbf{x}\in
C(\mathbf{r})}\Theta _{m}(d_{PM}(\mathbf{x},
\mathbf{r}_{d}))\}.
\end{equation}
The above modification is equivalent to saying that nodes which are further
than $m$ distance units from the destination are treated by the routing
algorithm \textit{as if they were exactly $m$ units away from the destination%
}. If a packet is at a node $\mathbf{r}$ such that all nodes directly linked
with $\mathbf{r}$ are further than $m$ units from its destination, then the
packet will be forwarded to a site selected randomly and uniformly from the
subset of $C(\mathbf{r})$ containing the nodes with the smallest queue size
in the set $C(\mathbf{r}).$ It can happen that the selected site can be
further away from the destination than the node $\mathbf{r}$.

Therefore, introduction of the\emph{\ cutoff parameter} $m$ adds more
randomness to the network dynamics. One could also say that the destination
attracts packets, but this attractive interaction has a finite range $m$:
packets further away than $m$ units from the destination are not being
attracted.

It is also possible to relate various values of the cutoff parameter
$m$ to different types of routing schemes used in real
packet-switching networks. Assume that each node $\mathbf{r}$
maintains a table containing all possible values of
$d_{PM}(\mathbf{x},\mathbf{r}_{d})$, for all possible destinations $%
\mathbf{r}_{d}$ and all nodes $\mathbf{x}\in C(\mathbf{r})$. Assume that
packets are routed according to this table by selecting nodes
minimizing distance, measured in the metric $d_{PM}$, traveled by a
packet from its origin to its destination. Such a routing scheme is
called \emph{table-driven routing} \cite{Saadawi94} and it is
equivalent to the routing algorithm $%
\mathbf{R}_{\infty }$. In this case, construction of the set $A_{\infty }(%
\mathbf{r})$ would require looking up appropriate entries in the stored
table.

Let us now define $D_{max}$ to be the largest possible distance
between two nodes in the network. When $m<D_{max}$, then for a given
${\mathbf{x}}$, we need to store values of
$d_{PM}({\mathbf{x}},{\mathbf{r}}_{d})$ only for nodes
${\mathbf{r}_{d}}$ which are less than $m$ units of distance away --
for all other nodes distance does not matter, since it will be treated
as $m$ by the routing algorithm. Hence, at each node $\mathbf{r}$ the
routing table to be stored is smaller than in the case when
$m=D_{max}$. The routing scheme based on this smaller routing table is
called the \emph{reduced table routing algorithm }\cite{Saadawi94} and
it is equivalent to the routing algorithm $\mathbf{R}_{m}$. In the
case when $m=D_{max}$ the routing algorithm $R_{m}=R_{\infty }.$

Finally, when $m=1$, the distances between hosts and destinations are
not considered in the routing process of packets. Therefore, there is
no need to store any table of possible paths at nodes of the network.
This case corresponds to the \emph{table-free routing algorithm
}\cite{Saadawi94} in which packets are routed randomly. Hence, this
algorithm can send packets on circuitous and long routes to their
destinations.

\section{Single packet delay}
One of the quantities characterizing the performance of a network is a
\emph{packet delay} $\tau_m$, frequently used in network performance literature
\cite{bert87,pac5,pac1,pac4,pac2,pac3,Stallings98}.
 In our case, the delay  will be defined as a
number of time steps elapsed from the creation of a packet to its
delivery to the destination address when the routing algorithm
$\mathbf{R}_{m}$ is used. In \cite{paper9} we found that the {\it free
packet delay}, or delay experienced by a packet when no other packets
are present, strongly determines behavior of the network, in
particular transition point to the congested state. Since in the case
of a single packet there is no interaction with other packets,
mathematical analysis of packet's dynamics is considerably simpler.
This analysis will be performed in what follows.

First of all, let us note that when
the routing algorithm $\mathbf{R}_{m}$ is used, and when the packet is
further than $m$ units away from its destination address, it performs
a random walk until it hits a node which is $m$ units away from the
destination, and then it follows the shortest path to the destination.
Obviously, several shortest paths might exists, so there is still
randomness in the packet's motion, but every time step its distance
from the destination decreases by one unit.

Let us denote by $\tau_m(\mathbf{r}_0, \mathbf{r}_d)$ the expected
delay time experienced by a packet which starts at $\mathbf{r}_0$ and
has destination address $\mathbf{r}_d$. For a lattice with periodic
boundary conditions, only relative position of $\mathbf{r}_0$ and
$\mathbf{r}_d$ is important. Therefore, we will choose $\mathbf{r}_d$
to be at the origin, and define $\tau_m(\mathbf{r}_0)=\tau_m(\mathbf{r}_0,
\mathbf{0})$.

From our discussion of the packet's motion we conclude that
$\tau_m(\mathbf{r}_0)$ is a sum of two parts:
\begin{equation} \label{splitdef}
\tau_m(\mathbf{r}_0)=\tau_{m,1}(\mathbf{r}_0) + \tau_{m,2}(\mathbf{r}_0),
\end{equation}
where $\tau_{m,1}(\mathbf{r}_0)$ is the expected time for a  random
walk to hit a node which is $m$ units away from the origin, and
$\tau_{m,2}(\mathbf{r}_0)$ is the expected time to reach the origin
starting from the node which is $m$ units away from the origin. We
will call $\tau_{m,1}(\mathbf{r}_0)$ a {\it random part}, and
$\tau_{m,2}(\mathbf{r}_0)$ a {\it semi-deterministic} part of the
delay $\tau_m(\mathbf{r}_0)$.

Obviously, for a single packet in the network
\begin{equation} \label{tau2}
\tau_{m,2}(\mathbf{r}_0) = \Theta _{m}(d_{PM}(\mathbf{r}_0,\mathbf{0})),
\end{equation}
and it is only $\tau_{m,1}(\mathbf{r}_0)$ that needs to be computed
(if $m<D_{max}$). It turns out that by modifying the problem slightly,
an analytical estimation of $\tau_{m,1}(\mathbf{r}_0)$ can be
obtained.

\subsection{Analytical estimation of the expected hitting time
for a random walk on a lattice $\L$.}
\label{analestim}

First, we observe that for a random walk which
start at $\mathbf{r}_0$, $\tau_{m,1}(\mathbf{r}_0)$ is the
expected time  of hitting the circle
 $S_m(0,d_{PM})=\{ \mathbf{r} \in \L:\
d_{PM}(\mathbf{r},\mathbf{0}) \leq m\}$

While the circle $S_m(0,d_{PM})$ defined in $d_{PM}$ metric is a
natural one to be used in our network model, it is not well suited
for the estimation of
$\tau_{m,1}(\mathbf{r}_0)$.
In order to carry such estimation, we will replace the circle $S_m(0,d_{PM})$
by the circle $S_m(0,d_{PE})$ in Euclidean metric, as explained below.

For any two points $\r_1 =(x_1,y_1)$ and $\r_2 =(x_2 ,y_2 )$ in $\L$
let us define the Euclidean distance with periodic boundaries between
this two points as
\bn
\de(\r_1,\r_2)=\sqrt{(\min\{x_1-x_2,L-(x_1-x_2)\})^2
+(\min\{y_1-y_2,L-(y_1-y_2)\})^2}.
\en
Notice that this metric is equivalent to the periodic Manhattan metric $d_{PM}$, in particular
\bn
\frac 1{\sqrt{2}}\,d_{PM}(\r_1,\r_2) \leq \de(\r_1,\r_2) \leq d_{PM}(\r_1,\r_2).
\en
For $\r\in\L$ let us set
\bn
\Vert \r \Vert=\de(\r,{\bf 0}).
\en
Hence, for any $a>0$, the circle of radius $a$ is the set
\bn
S_a=S_a(0,d_{PE})=\{\r\in\L:\ \Vert \r \Vert\leq a\}.
\en

Consider a simple random walk $\{X_k\}$, $k=0,1,2,\dots$ on $\L$.
Let $T_R(\r;L)$ be the
expected time of hitting the circle $S_R$ on a lattice $\L$
when the random walk $\{X_k\}$ starts at
$X_0=\r$.
\begin{thm}\label{tg}
Suppose that $R(1+\epsilon)<L/4$ and $R<\Vert \r\Vert<L/4$. If the
random walk $\{X_k\}$ starts at $\r$
then there exist a constant $C=C(\epsilon)>0$ such that
\bnn\label{lowbound}
 T_R(\r,L)\geq C L^2\log \left(\frac{\Vert\r\Vert}R\right)
\left[1+O\left(\frac 1L+\frac 1{R^2\log(\Vert \r\Vert/R)} \right) \right],
\enn
where we write $y(x)=O(x)$ whenever $\sup_{x>0} y(x)/x<\infty$.
\end{thm}
The proof of this theorem is based on the following lemma. Consider two
numbers $a$ and $c$ such that  $0<a<c\leq L/2$ and suppose that
$X_0=\r$ with $\Vert \r\Vert=b\in (a,c)$. Clearly, $S_a\subseteq S_c$,
$X_0\in S_c$ and $X_0\notin S_a$. Let $p_{a,c}(\r)$ be the
probability that the random walk $\{X_k\}$ will hit the circle $S_a$ before
exiting $S_c$.
\begin{lem}\label{lg}
If $f(\r)=\log(\Vert \r\Vert^2+1)$, then
\bn
 p_{a,c}(\r)\leq \frac{f(c)-f(b)}{f(c)-f(a)}
 =\frac{\log(c/b)+O(1/b^2)}{\log(c/a)+O(1/a^2)}.
\en
\end{lem}
{\it Proof of the Lemma.} The proof is conducted in the spirit of
\cite{Fayolle95}, the reader can also find in this book the
definition of submartingale and stopping time used further in this paper.

Observe that $\xi_k=f(X_k)$ is a submartingale with respect
to a filtration ${\cal F}_k=\sigma(X_0,X_1,\dots,X_k)$ generated by
the random walk $\{X_k\}$. Indeed, simple algebra shows that
\bn
\frac 14 \log((x+1)^2+y^2+1)&+&
\frac 14 \log((x-1)^2+y^2+1)+
\frac 14 \log(x^2+(y+1)^2+1)\\
&+&\frac 14 \log(x^2+(y-1)^2+1)>\log(x^2+y^2+1)
\en
and therefore
\bn
\E(\xi_{k+1}|{\cal F}_k)\geq \xi_k.
\en
Let the stopping time
\bn
\eta=\inf\{k>0:\ X_k\in S_a \mbox{ or } X_k\in \L\backslash S_c\}
\en
be the first time when the random walk leaves $S_c\backslash S_a$.
Then $\tilde{\xi}_k=\xi_{k\wedge\eta}$ is also a submartingale
\cite{Chow}, therefore
\bnn\label{submie}
\E \tilde{\xi}_k\geq \E \tilde{\xi}_0=f(b)
\enn
for all $k$. Obviously,  $\eta$ is finite a.s., so $\tilde{\xi}_k$
converges in $L^1$ to $\xi_{\eta}$ \cite{Chow}. On the other hand,
$f(X_{\eta})\leq f(a)$ if the random walk hits $S_a$ before
$\L\backslash S_c$ and $f(X_{\eta})\geq f(c)$ otherwise. Consequently,
\bn
\E [f(X_{\eta})\| X_{\eta}\in S_a]&\leq& f(a),\\
\E [f(X_{\eta})\| X_{\eta}\notin S_c]&\geq& f(c).
\en
Since $f(a)<f(b)$ and
\bn
\E(\xi_{\eta})=\E f(X_{\eta})
=\E [f(X_{\eta})\| X_{\eta}\in S_a] p_{a,c}(\r)+
 \E [f(X_{\eta})\| X_{\eta}\notin S_c] (1-p_{a,c}(\r)),
\en
the inequality (\ref{submie}) yields
\bn
p_{a,c}(\r)&\leq&\frac{\E [f(X_{\eta})\| X_{\eta}\notin S_c]-f(b)}
{\E [f(X_{\eta})\| X_{\eta}\notin S_c]-\E [f(X_{\eta})\| X_{\eta}\in S_a]}\\
&&\\
&\leq& \frac{\E [f(X_{\eta})\| X_{\eta}\notin S_c]-f(b)}
{\E [f(X_{\eta})\| X_{\eta}\notin S_c]-f(a)}\leq
\frac{f(c)-f(b)}{f(c)-f(a)}.
\en
Using the expansion
$\log(a^2+1)=2\log a +O(1/a)$ applied to $a$, $b$ and $c$
we conclude the proof of the Lemma.
$\Cox$
\vskip 5mm

\begin{figure}[htb]\label{figna}
\begin{center}
\includegraphics[scale=0.7]{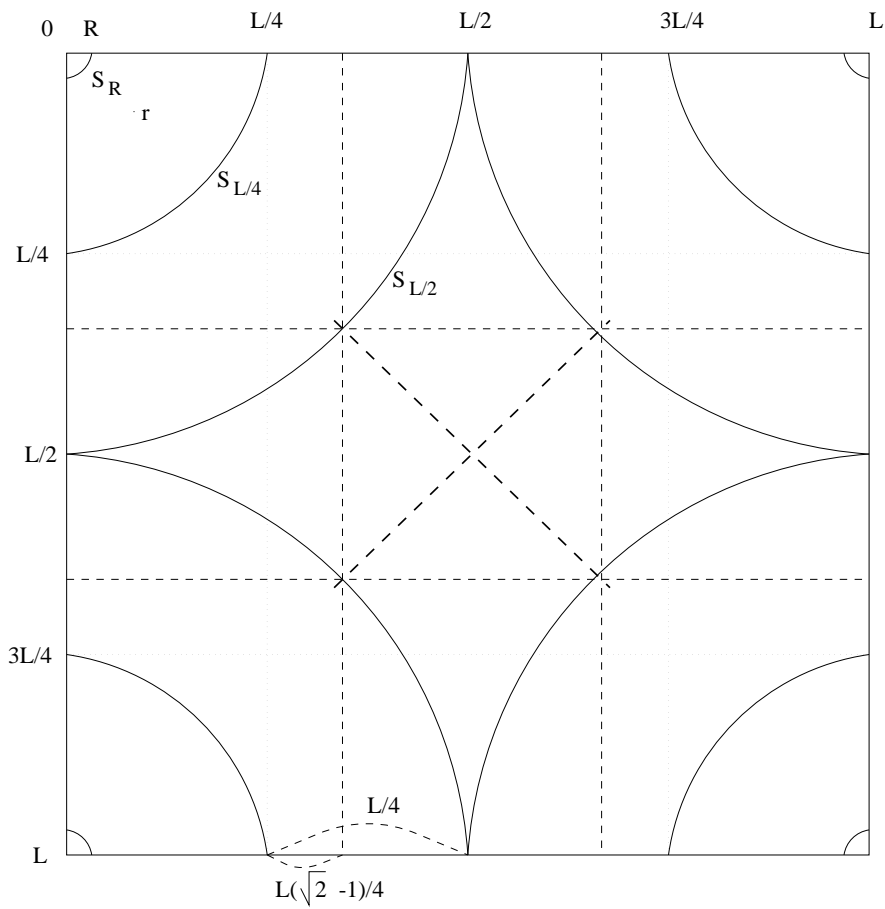}
\end{center}
\caption{Illustration to the proof of the Theorem~\ref{tg}}
\end{figure}

{\it Proof of Theorem~\ref{tg}.} The proof will proceed
in three steps. First, we will obtain the upper bound on the probability
of reaching $S_R$ prior to leaving $S_{L/2-1}$
when a random walk starts at a point $\r\in G$ where $G$
is the ring $S_{L/4}\backslash S_{L/4-1}$.
Next, we will estimate
the expected time of reaching $G$
starting from $\L\backslash S_{L/2-1}$.
In the second step, we will show that the expected time of
hitting $S_R$ when the random walk originates
inside $G$ is of order $L^2\log(L/2R)$. Finally, we will use the fact that
the expected time of hitting $S_R$ when the walk originates
at some $\r$ with $R<\Vert\r\Vert<L/4$ is at least as large
as the product of the probability of hitting $G$ prior to $S_R$
and the expected time of hitting $S_R$ starting from $G$.

{\bf Step 1.}

Let $G=S_{L/4}\backslash S_{L/4-1}$ be the set of lattice points
inside the ring of ``width'' one. Consider for  each $\r\in G$ a
simple random walk starting at $\r$, and a probability
$p_{R,L/2-1}(\r)$ that the random walk starting at $\r$ will hit $S_R$
before $\L\backslash S_{L/2-1}$. Let $p$ be smallest of these
probabilities, that is
\bnn\label{pmin}
p=\min_{\r\in G} p_{R,L/2-1}(\r),
\enn
then by Lemma~\ref{lg},
\bn
p\leq \frac{\log 2+O(1/L)}{\log(L/(2R))+O(1/L+1/R^2)}.
\en
Next, let us show that if the random walk starts in $\L\backslash
S_{L/2-1}$, then the minimum of all average times before hitting $G$ is
of order $L^2$. Indeed, when the random walk hits some $\r=(x,y)\in
G$, then $\de(\r)\leq L/4$ and therefore both $|x-L/2|\geq L/4$  and
$|y-L/2|\geq L/4$. However, for any $\r_1=(x_1,y_1)\notin S_{L/2-1}$
at least one of the values $x_1-L/2$ or $y_1-L/2$ lies inside the
segment $[-(\sqrt{2}-1)L/4-1,(\sqrt{2}-1)L/4+1]$ (see
Fig.~\ref{figna}). Consequently, the time in which the simple random
walk hits $G$ is stochastically larger\footnote{One random variable is
stochastically larger than another, if there is a probability space on
which both random variables are simultaneously defined and with
probability one the first one is at least as large as the other one.
For further references, see \cite{Chow}.} than $U$, the random variable
representing the time in
which one-dimensional simple random walk originating in $x\in
[-(2-\sqrt{2})L/4-1,(2-\sqrt{2})L/4+1]$ leaves the segment $[-\lfloor
L/4\rfloor,\lfloor L/4\rfloor]$\footnote{By $\lfloor a
\rfloor$ we mean the largest integer smaller than $a$.}. The expected
value of this random variable is known (see \cite{feller}) and equals
\bnn\label{expU}
(\lfloor L/4\rfloor)^2 -x^2\geq C_1 L^2
\enn
for some constant $C_1>0$, because $(2-\sqrt{2})/4<1/4$.

{\bf Step 2.} Let $\nu=\nu(\r)=\inf\{k:\ X_k\in S_R\}$ denote the
first time when the random walk starting at $X_0=\r\in G$ hits the
circle $S_R$. Consider a {\em stopped} random walk
$\tilde{X}_k=X_{k\wedge\nu}$ with $\tilde{X}_0=X_0$. Set $\eta_0=0$
and let
\bn
 \eta_n=\inf\{k>\eta_{n-1}:\ \tilde{X}_k\in G\mbox{ and }
  \tilde{X}_{k'}\notin S_{L/2-1}
 \mbox{ for some  }k'\in(\eta_{k-1},k)\}
\en
for $k=1,2,\dots$. Thus, $\eta_k$'s are consecutive times at which
$\tilde{X}_k$ finishes ``a loop'' from $G$ to $G$ visiting
$\L\backslash S_{L/2-1}$ for some time. Since the random walk eventually
hits $S_R$, only finitely many $\eta_k$'s will be defined.
 According to~(\ref{pmin}), the random number $N$ of such loops before
$X_k$ hits $\L\backslash S_{L/2-1}$ is stochastically larger than a
geometric  random variable $\bar{N}$ with parameter $p$ defined by
$\P(\bar{N}\geq n)= (1-p)^n$, $n=0,1,2\dots$. The probability that the
walk originating in $G$ will visit $\L\backslash S_{L/2}$ but will not
visit $S_R$, $n$ times in a row is at least $(1-p)^n$.
Consequently,
\bn
\eta_N=\sum_{i=1}^{N}(\eta_i-\eta_{i-1})\geq \sum_{i=1}^{N} U_i
\geq \sum_{i=1}^{\bar{N}} U_i,
\en
where $\{U_i\}$ is a sequence of random variables such that
$\E(U_i\|N)\geq C_1 L^2$ in accordance with~(\ref{expU}). Since
$\nu(\r)>\eta_N$, then for any $\r\in G$ we obtain
\bn
T_R(\r,L)&=&\E\nu(\r)>\E\eta_N\geq \sum_{n=1}^{\infty}
\E(\sum_{i=1}^{n} U_i\|
\bar{N}=n)\P(\bar{N}=n)\\ &\geq& C_1 L^2 \sum_{n=1}^{\infty} n
p(1-p)^{n-1}=\frac{C_1 L^2}{p}
\geq C_2 L^2 \log\frac{L}{2R}
\left[1+O\left(\frac{1}{L}+\frac{1}{R^2\log\frac L{2R}}\right)\right]
\en
where $C_2=C_1/\log 2$.

{\bf Step 3.}
Now suppose that $R<\Vert\r\Vert<L/4$. By Lemma~\ref{tg},
the event $A=\{X_k$ reaches $G$ before hitting $S_R\}$
has the probability
\bn
1-p_{R,L/4}(\r)\geq \frac{f(\Vert \r\Vert)-f(R)}{f(L/4)-f(R)}=
\frac{\log(\Vert \r\Vert/R)+O(1/R^2)
}{\log(L/(4R))+O(1/R^2)}:=q.
\en
Consequently,
\bn
T_R(\r,L)&=&\E \nu(\r)\geq \E(\nu(\r)\|A)\P(A)\geq q\min_{\r_1\in G}
\E\nu(\r_1)\\ &\geq& C L^2\log(\Vert \r\Vert/R)
\left[1+ O\left(\frac 1L+\frac 1{R^2\log(\Vert r\Vert/R)} \right) \right]
\en
since
\bn
\frac{\log(L/(4R))}{\log(L/(2R))}\geq
\frac{\log(1+\epsilon)}{\log(2+2\epsilon)}>0,
\en
and the Theorem is proven.
$\Cox$
\vskip 5mm

\begin{cor}
Under conditions of Theorem~\ref{tg}, if $R$ is fixed while both
$\Vert \r\Vert\to\infty$ and $L\to\infty$, then
\bn
 T_R(\r,L)\geq C L^2\log (\Vert\r\Vert/R)
[1+o(1)].
\en
\end{cor}

\subsection{The asymptotic behavior of $T_R$}
Here we will study the case when $L$ is so large, that a simple random walk
after appropriate rescaling is close to a Brownian motion  $B_t$ on a
square $\tL=[0,1]^2$ with periodic boundary conditions.\cite{Freedman}
Let $0<\ee<1$, $\r\in\tL$ and $\ttau_{\ee}(\r)$ be the expected time
in which Brownian motion starting from $\r$ will hit a circle of
radius $\ee$. To avoid a trivial answer, we always  assume that $\r$
lies outside of this circle. When the rescaled random walk starting at
$\r$ is close to the Brownian motion \cite{Freedman}, then for
sufficiently large $L$ and $R$
\bnn\label{limbeh}
 T_R(\r;L)\approx 2L^2 \ttau_{R/L}(\r/L).
\enn
Therefore, from bounds on  $\ttau_{\ee}(\r)$ we can deduce the
asymptotic behavior of $T_R(\r,L)$.

It follows from \cite{BASS}, p.109, that the function
$\ttau_{\ee}(\r)$ is a solution of the PDE on a square {\em with
periodic boundaries}
\bn
 \Delta\ttau&=&-2,\\
  \ttau(\r)|_{\r\in \partial C_{\ee}}&=&0,
\en
where for any $\ee>0$, $\partial C_{\ee}$ denotes  the boundary of a
circle of a radius $\ee>0$ around the origin ${\bf 0}$. Here we will not
be solving this PDE analytically. We will present estimates of
$\ttau$, which follow from a probabilistic nature of the model. The
following statement is essential, the idea of its proof comes from
\cite{DY}.
\begin{lem}
Consider a Brownian motion $B_t$ on a plane starting from $\r\in
\R^2$, such that $\rho=|\r|\in(a,b)$ and $0<a<b$. Let
$u=u(\rho;a,b)$ be the expected time until $B_t$ hits the circle
$C_a$, excluding the time spent outside the circle $C_b$, that is
\bn
u=\E \int_{0}^{\nu_a} 1_{\{|B_t|\leq b\}} dt
\en
where $\nu_a=\inf\{t:\ |B_t|\leq a\}$, then
\bnn\label{SPDEPS}
u(\rho;a,b)=b^2\log \frac{\rho}{a}-\frac{\rho^2-a^2}2.
\enn
\end{lem}
{\it Proof.} For a Brownian motion $B_t$ with $B_0=\r=(x,y)\in C_b$
and $u(\r)=\E\nu(\r)$ we define $\nu(\r)=\int_{0}^{\nu_a}
1_{\{|B_t|\leq b\}} dt$.
Consider a circle of a small radius $\rho_0$ around $\r$. Since
$u(\r)$ is a constant on $\partial C_b$, then from the symmetry of a circle and
by Markov Principle
\bn
 u(\r)&=&\frac 1{\phi_2-\phi_1}
\int_{\phi_1}^{\phi_2} u(x+\rho_0\cos\phi,y+\rho_0\sin\phi)d\phi\\
 &+&\frac 1{2\pi-\phi_2+\phi_1}
\int_{\phi_2}^{\phi_1+2\pi} u(\r)d\phi
+O({\rho_0}^2).
\en
In this equation the angles $\phi_1$ and $\phi_2$ are defined in such
a way that $\phi\in(\phi_1,\phi_2)$ corresponds to the points
$(x+\rho_0\cos\phi,y+\rho_0\sin\phi)$ lying inside the circle $C_b$
and $\phi\in(\phi_2,\phi_1+2\pi)$ corresponds to the points lying
outside of the circle $C_b$. Taking a Taylor expansion and letting
$\rho_0\to 0$ yields
\bnn\label{BCPDE}
\nabla u(\r)\cdot {\bf n}(\r)|_{\r\in \partial C_b}=0,
\enn
where ${\bf n}$ is a unit vector normal to $\partial C_b$ at $\r$.
On the other hand, for $\r$ lying inside the set $\{\r:\ |\r|<b\}$ we have
\bnn\label{PDEPS}
\Delta u=-2
\enn
(see \cite{DY}). Solving PDE~(\ref{PDEPS})
with the  boundary conditions~(\ref{BCPDE}) and
the condition $u(\r)|_{\r\in \partial C_a}=0$, we obtain~(\ref{SPDEPS}).
$\Cox$
\vskip5mm

Now, to get the desired estimates on $\ttau$, observe that
the geometry of the model implies
\bn
u\left(\rho\wedge\frac 12;\ee,\frac 12\right)  \leq \ttau_{\ee}(\r)
 \leq u\left(\rho;\ee,\frac 1{\sqrt{2}}\right)
\en
where $\rho$ is the distance from $\r$ to $\bf 0$ in Euclidean
periodic metric on $\tL$. In particular, using the R.H.S. of this
inequality we obtain the following result.
\begin{cor}
Whenever (\ref{limbeh}) takes place, $T_R(\r,L)$ is asymptotically
bounded from above by
\begin{equation}
 L^2\log\frac{\Vert\r\Vert}R  -(\Vert \r\Vert^2-R^2)+o(L^2).
 \label{upbound}
\end{equation}
\end{cor}
In terms of order, this equation matches closely the lower bound given
by (\ref{lowbound}). This is consistent with our results for the discrete
case and not really surprising, since the limit of a random walk is a Brownian
motion.

\subsection{Numerical results}
\begin{figure}
\begin{center}
a) \includegraphics{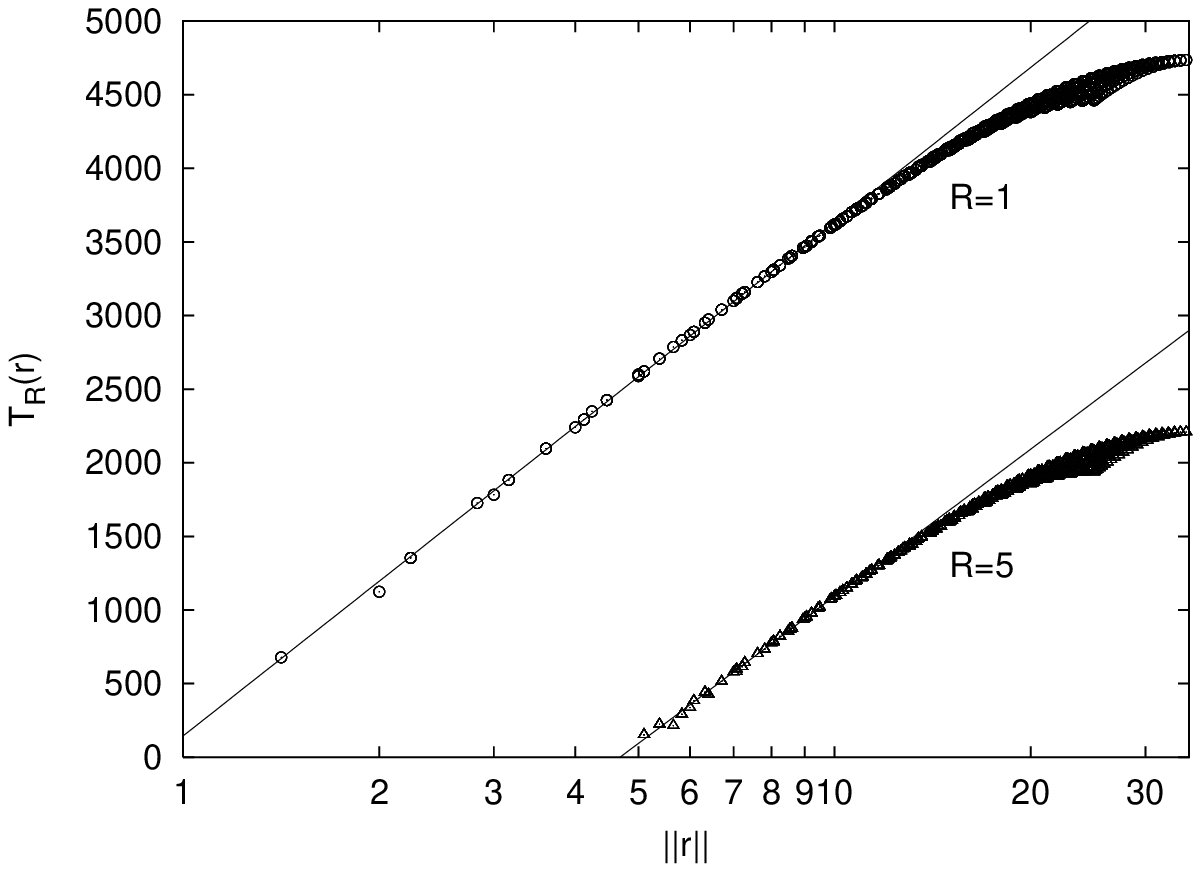}\\ b) \includegraphics{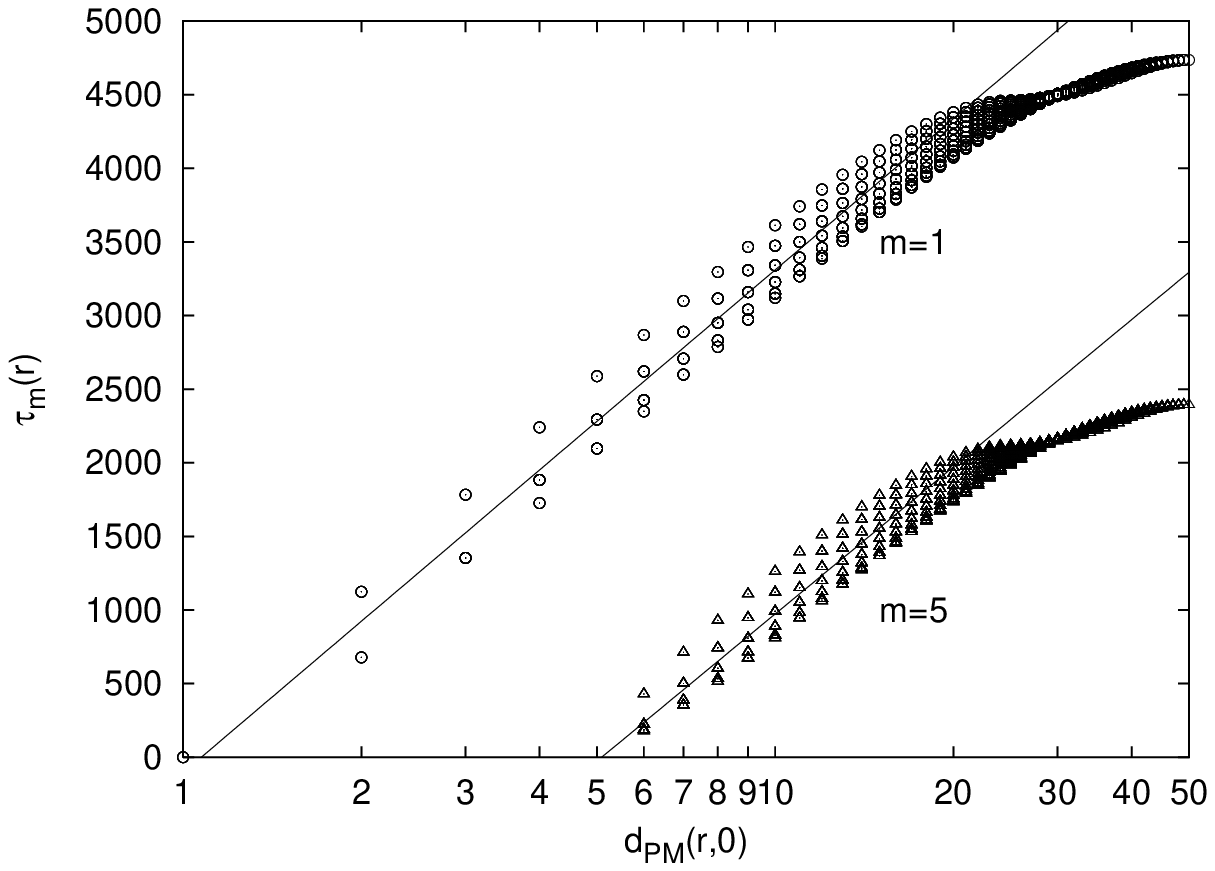}
\end{center}
\caption{Graphs of (a) $T_R(\mathbf{r},50)$ as a function of $||\mathbf{r}||$
for $R=1,5$ and (b) $\tau_m(\mathbf{r})$ as a function of
$d_{PM}(\mathbf{r}, \mathbf{0})$ for $m=1,5$ for a lattice $\L$ with
$L=50$. Continuous lines are the least square fits using points with
$||\mathbf{r}||\leq 10$.}
\label{fig200}
\end{figure}

In order to assess quality of analytical estimates of
$T_R(\mathbf{r})=T_R(\mathbf{r},L)$ obtained in the previous section,
we will compare them with values of $T_R(\mathbf{r})$ calculated
numerically by solving the system of linear equations
\begin{equation}
T_R(\mathbf{r})= 1+ \frac{1}{4}(
 T_R(\mathbf{r}+\mathbf{c}_x)+
T_R(\mathbf{r}-\mathbf{c}_x)+ T_R(\mathbf{r}+\mathbf{c}_y)+
T_R(\mathbf{r}-\mathbf{c}_y)),
\end{equation}
with periodic boundary conditions and $T_R(\mathbf{r})=0$ for every
$\mathbf{r} \in \L$ such that $d_{PE}(\mathbf{r},\mathbf{0})
\leq R$. Figure \ref{fig200}a is a semi-log plot of $T_R(\mathbf{r})$
as a function of $||\mathbf{r}||$ for the lattice $L\times L=50 \times
50$ and two values of $R$, $R=1$ and $R=5$. Each lattice node for
which $||\mathbf{r}||>R$ is represented by a single point on the
graph. One can clearly see that for $||\mathbf{r}||$ smaller than
about 10, these points form a straight line, in agreement with
estimations (\ref{lowbound}) and (\ref{upbound}).

Once we notice that for every $\mathbf{r} \in \L$ such that
$d_{PM}(\mathbf{r},\mathbf{0})=m$ we have
\begin{equation}
 \frac{m\sqrt{2}}{2} \leq d_{PE}(\mathbf{r},\mathbf{0}) \leq m,
\end{equation}
we can obtain the following bounds on $\tau_{m,1}(\mathbf{r})$:
\begin{equation}
\label{estimT}
T_m(\mathbf{r}) \leq \tau_{m,1}(\mathbf{r}) \leq T_{m
\sqrt{2}/2}(\mathbf{r}).
\end{equation}
The above relationship is well illustrated in Figure \ref{fig200}b,
which shows a graph of $\tau_{m,1}(\mathbf{r})$ as a function of
$d_{PM}(\mathbf{r},\mathbf{0})$ for $L=50$ and $m=1,5$. As before, the
values of $\tau_{m,1}(\mathbf{r})$ were obtained by solving the system
of linear equations
\begin{equation}
\tau_{m,1}(\mathbf{r})= 1+ \frac{1}{4}(
 \tau_{m,1}(\mathbf{r}+\mathbf{c}_x)+
\tau_{m,1}(\mathbf{r}-\mathbf{c}_x)+ \tau_{m,1}(\mathbf{r}+\mathbf{c}_y)+
\tau_{m,1}(\mathbf{r}-\mathbf{c}_y)),
\end{equation}
with  periodic boundary conditions and $\tau_{m,1}(\mathbf{r})=0$ for
every $\mathbf{r}
\in
\L$ such that $d_{PM}(\mathbf{r},\mathbf{0})
\leq m$.
In  the aforementioned figure, the points close to the origin do not
lie on a straight line, but lie in an area bounded by two straight
lines, as expected from (\ref{estimT}).

\section{Average delay}
In a network model investigated in \cite{paper9}, packets were created
at each node with a destination address randomly selected among all
nodes of the lattice. A useful quantity characterizing delay experienced
by packets under such circumstances is an average delay
$\overline{\tau}_m$, defined as
\begin{equation}
\overline{\tau}_m=\frac{1}{L^2} \sum_{\mathbf{r} \in \L} \tau_m(\mathbf{r}).
\end{equation}
Similarly as in (\ref{splitdef}), we can write the average delay
$\overline{\tau}_m$ as a sum of the average random and the average semi-deterministic
parts, denoted by $\overline{\tau}_{m,1}$ and
$\overline{\tau}_{m,2}$, respectively.

Using (\ref{tau2}), we will calculate the average semi-deterministic
part of the average delay. First, let us define $N(k)$ to be a number
of sites $\mathbf{r}\in \L$ such that
$d_{PM}(\mathbf{r},\mathbf{0})=k$, $0
\leq k  \leq L$. Then we can write $\overline{\tau}_{m,2}$ as
\begin{equation} \label{avtau2}
\overline{\tau}_{m,2}=\frac{1}{L^2} \sum_{\mathbf{r} \in \L} \tau_{m,2}(\mathbf{r})=
\frac{1}{L^2} \sum_{k=0}^{L} N(k) \Theta _{m}(k).
\end{equation}
For simplicity,  and without much loss of generality, in what follows
we will assume that $L$ is even. It is straightforward to establish
that for even $L$
\begin{equation}
N(k)= \left\{ \begin{array}{ll}
 1     &  \mbox{if $k=0$} \\
 4k    &  \mbox{if $0<k<L/2$} \\
 2L-2  &  \mbox{if $k=L/2$} \\
 4(L-k)&  \mbox{if $L/2 <k <L$} \\
 1     &  \mbox{if $k=L$}
 \end{array}
 \right.
\end{equation}
which can written in a more compact form as
\begin{equation}
N(k) = \delta_{0,k} + \delta_{L,k} - 2\delta_{L/2,k} +2L - |4k-2L|,
\end{equation}
where $\delta_{i,j}=1$ if $i=j$ and $\delta_{i,j}=0$ otherwise. Using
this result and computing the sum in (\ref{avtau2}) we obtain
\begin{equation} \label{tau2final}
\overline{\tau}_{m,2}= \left\{ \begin{array}{ll}
 {\displaystyle m-\frac{2m^3+m}{3L^2}},     &  \mbox{if $m<L/2$} \vspace{.2cm}\\
 {\displaystyle \frac{L}{2} -\frac{2(L-m)^3 + L-m}{3L^2}},     &  \mbox{otherwise.}
 \end{array}
 \right.
\end{equation}

\begin{figure}
 \begin{center}
  \includegraphics{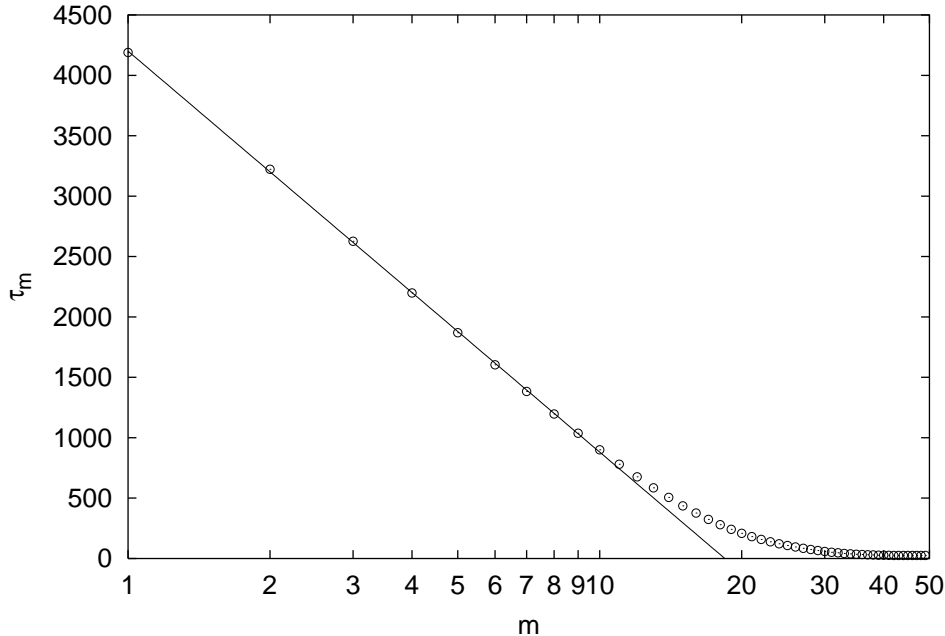}
 \end{center}
\caption{Average delay $\overline{\tau}_m$ of a free packet as a function of
$m$ for a periodic lattice $50 \times 50$. The continuous line
represents the least squares fit to the first 10 points.}
\label{fig100}
\end{figure}

Since the average semi-deterministic part of the average delay is
always smaller than $m$, for small $m$ it will be negligible compared
to the random part. Therefore, in the small $m$ regime, we can expect
that the leading term in $\overline{\tau}_m$ is  a linear function of
$\log(m)$, according to our analytical estimate from the previous
section. Figure \ref{fig100} shows that it is indeed the case, as
illustrated for $L=100$.
An important observation which can be made from this figure is that
$\overline{\tau}_m$ stays close to its $m=L$ value
($\overline{\tau}_L$=L/2, see eq. \ref{tau2final}) when $m$ is close
to $L$. This means that making $m$ slightly smaller than $L$ does not
increase delay significantly.

\section{Network scalability}
Every network at some point of its life span needs to be expanded. It
is obvious that as the number of nodes increases, the average delay
increases as well, since the number of links to be traversed by a
given packet becomes larger. However, the increase in delay,
is not the only problem encountered when the network expands. Each
node $\mathbf{r}$ stores a routing table, which in our model
contains routing information for all nodes $\mathbf{x} \in \L$ such
that $d_{PM}(\mathbf{r},\mathbf{x}) \leq m$. If by $M(m)$ we denote
the number of nodes which are up to $m$ links away from a given node,
we can say that the memory required to store the routing table is
proportional to $M(m)$, which can be readily computed:
\begin{equation} \label{memory}
M(m)=\sum_{k=1}^{m} N(k) =\left\{ \begin{array}{ll}
 2m(m+1),     &  \mbox{if $0<m<L/2$} \vspace{.2cm}\\
 L^2-2(L-m)(L-m-1)-2    ,     &  \mbox{if $L/2 \leq m < L$}.
 \end{array}
 \right.
\end{equation}
Let us now assume that the ``cost'' of operating of a single node
with routing algorithm $\mathbf{R}_m$ is given by
\begin{equation}
c(m,a)= \overline{\tau}_m + a M(m),
\end{equation}
where $a$ is a nonnegative parameter describing the relative cost of
memory vs. average delay.
This cost function has been introduced to investigate strategies
which could minimize both average delay and memory storage requirements
at a node. The above form of $c(m,a)$ simply means that the
cost is a linear combination of memory used to store the routing
table and the average delay experienced by packets. By using this
form we want to express the fact that the delay experienced by
packets decreases utility of the network, and therefore increases its
``cost''.
\begin{figure}
\begin{flushleft}
\includegraphics[scale=0.8]{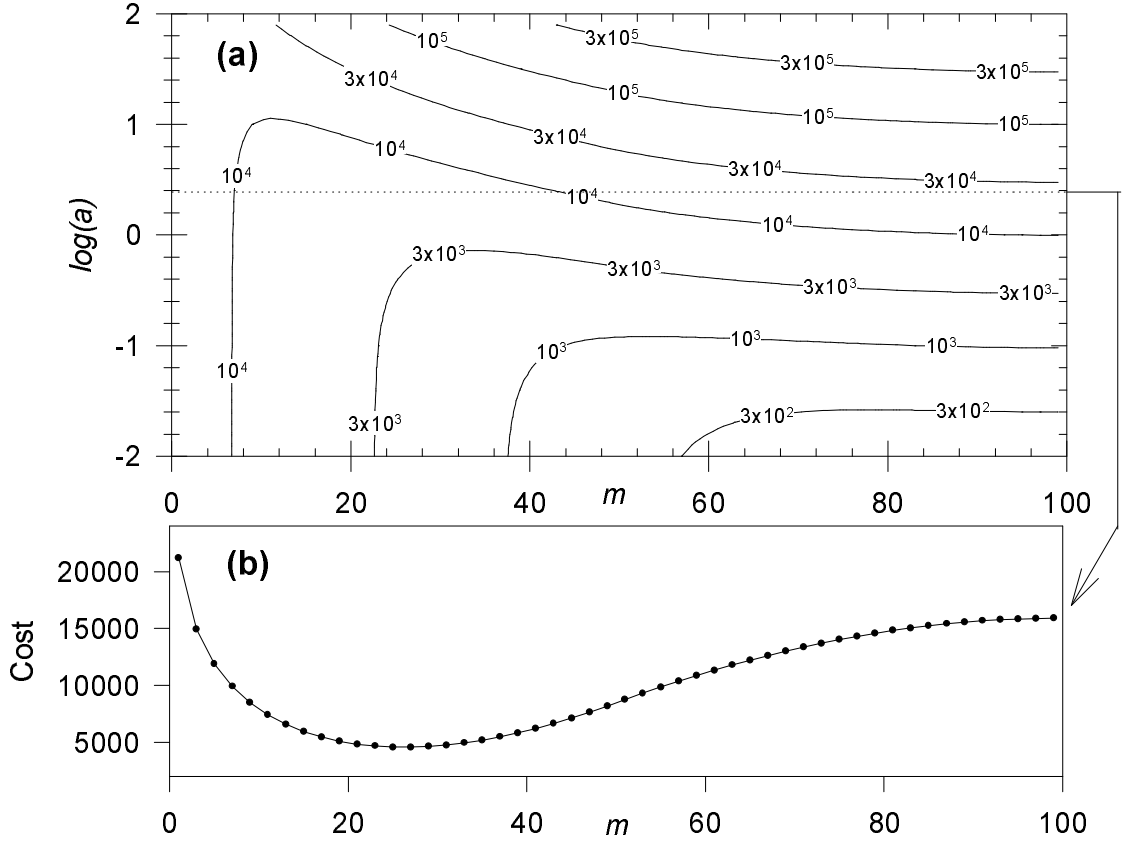}
\end{flushleft}
\caption{Cost function as a function of $a$ and $m$ shown as
 a contour plot (a). Part (b) shows the graph of the cost
function as a function of $m$ for a fixed value of $a$ ($a=1.58$).}
\label{costfig}
\end{figure}

Figure \ref{costfig} shows how the total cost $c(m,a)$ depends on $m$
and $a$ for $L=50$. For any given value of $a$, one can find the value
of $m$ which minimizes the total cost, as shown in Figure
\ref{costfig}b.

Obviously, when $a$ is very small, i.e., when the cost of storage is
negligible, the total cost is minimal at $m=L$. This means that if the
delay alone is taken into consideration, full table routing is always
a best choice. In that case, $c(m,a)$ will increase with $L$ as $L^2$,
meaning that the cost per node will grow proportionally to the number
of nodes in the network.

When $a$ is large,  the situation is very different. Let us assume,
for example, that the value of  $a$ is large enough so that the value
of $m$ minimizing $c(m,a)$ is small compared to $L$. In this case, the
random part of $\overline{\tau}_m$ is much larger than the
semi-deterministic part, and we can assume that the leading term of
$\overline{\tau}_{m}$ has the form
\begin{equation}
\overline{\tau}_{m} \approx \overline{\tau}_{1,m} =
AL^2\log{\frac{BL}{m}},
\end{equation}
where $A$ and $B$ are constants independent of $L$, and therefore
\begin{equation}
c(m,a) \approx AL^2\log{\frac{BL}{m}} + 2am(m+1).
\end{equation}
The above cost function is minimized by
\begin{equation}
m=\frac{\sqrt{a^2+4aAL^2}}{4a} -\frac{1}{2},
\end{equation}
which is  an asymptotically linear function of $L$. This means the
optimal strategy which should be used to minimize the ``cost'' of
the network is to increase $m$ proportionally to $L$, or in other
words, to increase the size of the routing table proportionally to
the number of nodes in the network. Note that in this case the cost will
still grow with $L$, and for large values of $L$ it will grow like $L^2$,
similarly as in the case
of very small $a$.

\section{Conclusion}

We have investigated individual packet delay in a model of data networks
with table-free, partial table and full table routing. We presented
analytical estimates for the average packet delay in a network with
small partial routing table and compared them with numerical results.
We have also examined the dependence of the delay on the size of a
network and on the size of a partial routing table. Assuming the
total ``cost'' of a network with routing algorithm $\mathbf{R}_m$ is a
linear combination of memory used to store the routing table and the
average delay experienced by packets, we discussed consequences of our
findings for network scalability. If we are concerned primary with the
speed of the network and the memory cost is not important, full table
routing is the best choice. On the other hand, if the primary factor
influencing the total cost is an amount of memory used to store
routing tables, the optimal strategy which should be used to
minimize the cost is to keep a size of a routing table
proportional to a number of nodes in a network. In that case, the
cost per node $c(m,a)$ grows linearly with the size of the network.

\subsection*{Acknowledgements}
The authors acknowledge partial financial support from the Natural
Sciences and Engineering Research Council (NSERC) of Canada and The
Fields Institute for Research in Mathematical Sciences. The
discussion of the problem analyzed in Section~\ref{analestim} with
Mikhail Menshikov was very helpful. One of the authors (H.F.)
expresses gratitude to the Department of Mathematics and Statistics,
University of Guelph, for hosting him as an NSERC Postdoctoral
Fellow.

\end{document}